\documentclass[a4paper]{article}

\usepackage{dingbat}

\usepackage{calc}  
\usepackage{enumitem}
\usepackage{hyperref}
\makeatletter
\def\namedlabel#1#2{\begingroup
    #2%
    \def\@currentlabel{#2}%
    \phantomsection\label{#1}\endgroup
}
\makeatother

\usepackage{amssymb}
\usepackage{amsmath}
\usepackage{proof}

\usepackage{tikz}
\usetikzlibrary{arrows,automata}

\usepackage{xspace}
\newcommand{\cA}{\ensuremath{\mathcal{A}}\xspace}

\newcommand{\cD}{\ensuremath{\mathcal{D}}\xspace}
\newcommand{\cF}{\ensuremath{\mathcal{F}}\xspace}
\newcommand{\cG}{\ensuremath{\mathcal{G}}\xspace}

\newcommand{\tx}[1]{\relax
  \ifmmode\mathchoice
    {\hbox{\the\textfont0\relax#1}}%
    {\hbox{\the\textfont0\relax#1}}%
    {\hbox{\the\scriptfont0\relax#1}}%
    {\hbox{\the\scriptscriptfont0\relax#1}}%
  \else{\relax#1}\fi}

\newcommand{\SF}[1]        {\tx{\sf #1}}

\newcommand {\limp}        {\mathbin{\Rightarrow}}
\newcommand {\leqv}        {\mathbin{\Leftrightarrow}}
\newcommand {\lconj}       {\mathbin{\wedge}}

\newcommand {\lneg}        {\mathop{\neg}}

\newcommand{\comm}[2]      {{\copyright_{#1}[#2]}}
\newcommand{\gf}[2]        {{@_{#1}[#2]}}

\newcommand {\nxt}         {\mathop{\SF{X}}}
\newcommand {\sometime}    {\mathop{\SF{F}}}
\newcommand {\always}      {\mathop{\SF{G}}}

\newcommand{\trans}                 {{\longrightarrow}}
\newcommand{\transact}[1]           {\stackrel{#1}{\trans}}

\newcommand{\sat}     {\Vdash}

\ifx \theorem \undefined
\newtheorem{definition}{\vspace{1mm}Definition}[section]

\newtheorem{example}[definition]{\vspace{1mm}Example}
\newtheorem{lemma}[definition]{\vspace{1mm}Lemma}
\newtheorem{theorem}[definition]{\vspace{1mm}Theorem}

\newtheorem{proposition}[definition]{\vspace{1mm}Proposition}

\fi

\newcommand{\qurtainsproof}    	{\hfill{${\Box}$}\par \vspace{1ex}}

\newcommand{\proof}[1]         	{\noindent{\textsl{Proof: }{#1}}\qurtainsproof\mbox{}\\[-5mm]}

\newcommand{\DTL}              	{\textrm{DTL}}

\newcommand{\LL}               	{{\mathcal L}}

\newcommand{\Id}               	{\textit{Id\/}}
\newcommand{\Prop}             	{\textit{Prop\/}}
\newcommand{\Ev}               	{\textit{Ev\/}}
\newcommand{\last}             	{\textit{last\/}}

\newcommand{\Ids}              	{\textit{Ids\/}}

\newcommand{\Mod}			   	{\textrm{Mod}}
\newcommand{\cL}			   	{{\mathcal L}}

\newcommand{\subf}             	{\textit{subf}}

\newcommand{\closure}    		{\textit{closure}}
\newcommand{\lit}             	{\textrm{\it Lit}}
\newcommand{\Val}             	{{\mathcal V}}

\newcommand{\da}               	{\!\!\downarrow}

\newcommand{\val}              	{\vartheta}

\newcommand{\nats}             	{{\mathbb N}}

\newcommand{\tuple}[1]         	{\langle #1 \rangle}

\newcommand{\hide}[1] {}

\begin{document}

\title{\sc  B\"uchi automata for distributed temporal logic}

\author{Jaime Ramos \\
\footnotesize{Dep. Matem\'atica, Instituto Superior T\'ecnico, Universidade de Lisboa, Portugal}\\
\footnotesize{SQIG, Instituto de Telecomunica\c{c}\~oes, Portugal} \\
\footnotesize{\tt jaime.ramos@tecnico.ulisboa.pt}\\[3mm]
}


\maketitle

\begin{abstract}
The distributed temporal logic DTL is a logic for reasoning about temporal
properties of distributed systems from the local point of view of the
system's agents, which are assumed to execute sequentially and to interact
by means of synchronous event sharing. Different versions of DTL have been
provided over the years for a number of different applications, reflecting
different perspectives on how non-local information can be accessed by
each agent. In this paper, we propose a  novel notion of distributed 
B\"uchi automaton envisaged to encompass DTL with 
a model-checking mechanism.\\[2mm]
\end{abstract}

\section{Introduction}

The distributed temporal logic DTL was introduced in~\cite{hde:ccal:98} as
a logic for specifying and reasoning about distributed information
systems. DTL allows one to reason about temporal properties of distributed
systems from the local point of view of the system's agents, which are
assumed to execute sequentially and to interact by means of synchronous
event sharing. In DTL, distribution is implicit and properties of entire
systems are formulated in terms of the local properties of the agents and
their interaction. The logic was shown to be decidable, as well as trace-consistent, which makes it suitable for model-checking tasks.

Different versions of distributed temporal logic have been given over the years for a number of
different applications, reflecting different perspectives on how non-local
information can be accessed by each agent. In particular, DTL has proved
to be useful in the context of security protocol analysis in order to
reason about the interplay between protocol models and security
properties~\cite{ccal:vig:bas:04b,ccal:vig:bas:04c,bas:ccal:jabr:vig:10}.
However, most of the results for security protocol analysis and for other case
studies were obtained directly by semantic arguments.\footnote{DTL is
closely related to the family of temporal logics whose semantics are based
on the models of true concurrency introduced and developed
in~\cite{lod:ram:thi:92,lod:thi:87,ram:96}. In particular, the semantics
of these logics are based on a conflict-free version of Winskel's event
structures~\cite{win:87}, enriched with information about sequential
agents.} To overcome this problem, a 
labeled tableaux system for DTL was proposed
in~\cite{bas:ccal:jabr:vig:07,bas:ccal:jabr:vig:09}. The main goal was to
have a usable deductive system in which deductions followed closely
semantic arguments, also thanks to the labeling of the formulas along with
a labeling algebra capturing the different semantic properties.

The labeled tableaux system was proved to be sound and complete, but decidability was not considered
in~\cite{bas:ccal:jabr:vig:07,bas:ccal:jabr:vig:09} and the system
included an infinite closure rule to capture eventualities that are always
delayed. Hence, the labeled system proved to be quite hard to use in
practice although several properties can still be proved using only the
tableaux system. For instance, the correctness of the \emph{two-phase
commit protocol} is one of such examples where a decision procedure is
not needed. The DTL specification for a simplified version of the protocol
as well as a proof of correctness using labelled tableaux can be found
in~\cite{bas:ccal:jabr:vig:09}.

Nevertheless, DTL was shown to be decidable via a translation to linear
temporal logic (LTL). However, when translating DTL specifications into
LTL specifications, we lose one of the main advantages of DTL, namely the
naturalness of the distributed nature of DTL, which allows for more
natural and simpler specifications. Later, in~\cite{ccal:mpg:jabr:lv:17}, a
\textit{decidable tableaux system} was proposed for DTL. The tableaux system was
built on top of a tableaux system for LTL as
presented in~\cite{kroger:merz:08}. Similar systems for LTL have also been
proposed, e.g.,~\cite{lic:pnu:00}. In the case of DTL, the tableaux system integrated
in a smooth way both the usual rules
for the temporal operators and rules for tackling the specific
communication features of DTL. 

In this paper, we take a first step towards empowering DTL with model-checking 
tools. 
Nowadays, systems are becoming more and more complex which makes the task of verification
such systems harder. Model-checking stands out as a tool well suited for automatic verification, 
which has been successfully used in industry with several well documented cases~\cite{mcm:93,hol:04,bar:ram:16}. 
Depending on the  temporal logic considered~\cite{pnu:77,cla:eme:82}, the approach to model-checking is different\cite{eme:cla:80,cla:eme:sis:86, var:wol:94}.
In the case of DTL, we adopt an approach closer to the usual approaches in LTL, 
based on B\"uchi automata~\cite{var:wol:94}. Our goal is to use B\"uchi automata to capture DTL models.

For the local component of our automata, we follow closely the ideas in~\cite{var:wol:94,bai:kat:08}.
It is worth mentioning that, similar to~\cite{bai:kat:08}, in which an anchored
version of LTL is considered, in this paper we consider an \textit{anchored
version of DTL}. This anchored version of DTL is less
expressive in terms of global reasoning since DTL does not include
global temporal operators and, thus, we cannot use the usual correspondence between
anchored and floating semantics of temporal logic. However, it let us focus on the distributed 
nature and synchronization primitives of the logic.

We proceed as follows. In Section~\ref{sec:dtl}, we briefly introduce DTL, its syntax, semantics and 
some auxiliary notions that will be useful later. In Section~\ref{sec:dba}, we present distributed B\"uchi automata for DTL and 
prove the correctness of the construction with respect to the semantics of DTL. In Section~\ref{sec:conc}, we conclude and discuss future work.

\section{The Distributed Temporal Logic \texorpdfstring{$\DTL_\emptyset$}{DTL}}\label{sec:dtl}

As we mentioned above, a number of variants of $\DTL$ have been considered
in the past, especially to adapt it to specific applications and case
studies. In this paper, we consider an anchored variant of $\DTL$ that we
call $\DTL_\emptyset$ and that has the following syntax and semantics.

\subsection{Syntax}

The logic is defined over a \emph{distributed signature}
\begin{displaymath}
\Sigma=\tuple{\Id,\{\Prop\}_{i\in\Id}}\,,
\end{displaymath}
where $\Id$ is a finite non-empty set (of \emph{agent identifiers}) and, for each agent $i\in\Id$, $\Prop_i$ is a set of \emph{local state propositions}, which, intuitively characterize the current local states of the agents. We assume that $\Prop_i\cap\Prop_j=\emptyset$, for $i\neq j$.

The \emph{local language} $\LL_i$ of each agent $i\in\Id$ is defined by
\begin{displaymath}
\LL_i ::= \Prop_i \mid \lneg \LL_i \mid \LL_i \limp \LL_i \mid 
          \nxt \LL_i \mid \always\LL_i \mid \comm{j}{\LL_j}
\end{displaymath}
with $j\in\Id$. We will denote such \emph{local formulas} by the letters
$\varphi$ and $\psi$. As the names suggests, local formulas hold locally
for the different agents. For instance, locally for an agent $i$, the
operators $\nxt$ and $\always$ are the usual \emph{next (tomorrow)} and
\emph{always in the future} temporal operators, whereas the
\emph{communication formula} $\copyright_j[\psi]$ means that agent $i$ has
just communicated (synchronized) with agent $j$, for whom $\psi$ held.

Other logical connectives (conjunction $\lconj$, disjunction $\vee$, true
$\top$, etc.) and temporal operators (sometime in
the future $\sometime$) can be
defined as abbreviations as is standard.

The \emph{global language} $\LL$ is defined by
\begin{displaymath}
\LL ::= \gf{i}{\LL_i} \mid \lneg\LL \mid \LL\limp \LL
\end{displaymath}
with $i\in\Id$. We will denote the \emph{global formulas} by $\alpha$, $\beta$ and $\delta$. A \emph{global formula} $@_i[\varphi]$ means that the local formula $\varphi$ holds for agent $i$.

In the sequel, we will need some auxiliary notions. The set of $i$-{\it literals} is the set of all state propositions and their negations:
$$
\lit_i=\Prop_i\cup\{\lneg p\mid p\in\Prop_i\}.
$$
An $i$-valuation $v$ is a set of $i$-literals such that for each $p\in\Prop_i$, $p\in v$ iff $\lneg p\notin v$. The set of all $i$-valuations is denoted by $\Val_i$. Observe that $\Val_i\subseteq 2^{\lit_i}$.

Given $i\in\Id$ and $\varphi,\psi\in \LL_i$, we say that $\psi$ is an $i$-{\it subformula} of $\varphi$ if $\psi$ is $\varphi$ or:
\begin{itemize}

\item $\varphi$ is $\lneg\varphi_1$ and $\psi$ is  a subformula of $\varphi_1$;

\item $\varphi$ is $\varphi_1\limp\varphi_2$ and $\psi$ is  a subformula of $\varphi_1$ or of  $\varphi_2$;

\item $\varphi$ is $\nxt\varphi_1$ and $\psi$ is  a subformula of $\varphi_1$;

\item $\varphi$ is $\always\varphi_1$ and $\psi$ is  a subformula of $\varphi_1$.

\end{itemize}
We denote by $\subf_i(\varphi)$ the set of all $i$-subformulas of $\varphi$. When no confusion arises, we drop the reference to $i$ and talk about subformulas. For instance, the set of subformulas of 
$\always (p\limp\comm{j}{q_1\limp q_2})\in\LL_i$, 
$\subf_i(\always (p\limp\comm{j}{q_1\limp q_2}))$, is
$$
\{\always (p\limp\comm{j}{q_1\limp q_2}), p\limp\comm{j}{q_1\limp q_2}, p, \comm{j}{q_1\limp q_2}\}.
$$
Note that, from the point of view of agent $i$, formula $\comm{j}{q_1\limp q_2}$ has no further structure and is treated as atomic.

The $i$-{\it closure} of $\varphi$ is the set of all its subformulas and their negations with the proviso that $\lneg\lneg\psi$ is identified with $\psi$, that is, 
$$
\closure_i(\varphi)=\subf_i(\varphi)\cup\{\lneg\psi\mid\psi\in\subf_i(\varphi)\}.
$$
Again, when no confusion arises, we will talk about the closure of a formula.

We also define similar concepts for the global language. Given $\alpha,\beta\in\LL$, we say that $\beta$ is a \textit{subformula} of $\alpha$ if 
$\beta$ is $\alpha$ or:
\begin{itemize}
\item $\alpha$ is $\gf{i}{\psi}$ and $\beta\in\subf_i(\varphi)$
\item $\alpha$ is $\lneg\alpha_1$ and $\beta$ is a subformula of $\alpha_1$;
\item $\alpha$ is $\alpha_1\limp\alpha_2$ and $\beta$ is a subformula of $\alpha_1$ or of $\alpha_2$.
\end{itemize}
For instance, the set of subformulas of  formula
$\gf{i}{\nxt(p\limp\comm{j}{q})}\limp\gf{j}{\nxt q}$, $\subf(\gf{i}{\nxt(p\limp\comm{j}{q})}\limp\gf{j}{\nxt q})$, is:
$$\begin{array}{l}
		\{\gf{i}{\nxt(p\limp\comm{j}{q})}\limp\gf{j}{\nxt q},\gf{i}{\nxt(p\limp\comm{j}{q})},\gf{j}{\nxt q}\}\\[2mm]
		\qquad\qquad\qquad\qquad\qquad\qquad{}\cup\subf_i(\nxt(p\limp\comm{j}{q}))\cup\subf_j(\nxt q)=\\[2mm]
		\{\gf{i}{\nxt(p\limp\comm{j}{q})}\limp\gf{j}{\nxt q},\gf{i}{\nxt(p\limp\comm{j}{q})},\gf{j}{\nxt q}\}\\[2mm]
		\qquad\qquad\qquad\qquad\qquad\qquad{}\cup\{\nxt(p\limp\comm{j}{q}),p\limp\comm{j}{q},p,\comm{j}{q},\nxt q,q\}.
\end{array}$$

The \textit{closure} of  $\alpha$ is the set of all its subformulas and their negations with the proviso that $\lneg\lneg\beta$ is identified with $\beta$, that is, 
$$
\closure(\alpha)=\subf(\alpha)\cup\{\lneg\beta\mid\beta\in\subf(\alpha)\}.
$$

Finally, given a set $B\subseteq\closure(\alpha)$ and $i\in\Id$ we denote $B\da_i$ the subset of $B$ that contains 
all the global formulas of $B$ and no local formulas other than those of agent $i$, that is, $B\da_i$ satisfies the following conditions:
\begin{itemize}
\item $B\da_i\subseteq B$;
\item $B\da_i\cap\LL=B\cap\LL$;
\item $B\da_i\cap\LL_i=B\cap\LL_i$;
\item $B\da_i\cap\LL_j=\emptyset$, for $j\neq i$.
\end{itemize}
For instance, if $B=\{\gf{i}{\nxt(p\limp\comm{j}{q})}\limp\gf{j}{\nxt q},\nxt(p\limp\comm{j}{q}),\comm{j}{q},\nxt q\}$ then
$$
B\da_i=\{\gf{i}{\nxt(p\limp\comm{j}{q})}\limp\gf{j}{\nxt q},\nxt(p\limp\comm{j}{q}),\comm{j}{q}\}
$$
and 
$$
B\da_j=\{\gf{i}{\nxt(p\limp\comm{j}{q})}\limp\gf{j}{\nxt q},\nxt q\}.
$$

\subsection{Semantics}

The interpretation structures of $\LL$ are labeled distributed
life-cycles, built upon a simplified form of Winskel's \emph{event
structures}~\cite{win:87}. 

A \emph{local life-cycle} of an agent $i \in \Id$ is a countable infinite,
discrete, and well-founded total order
\hbox{$\lambda_i=\tuple{\Ev_i,\leq_i}$}, where $\Ev_i$ is the set of {\em
local events} and $\leq_i$ the \emph{local order of causality}. We define
the corresponding \emph{local successor relation}
$\rightarrow_i\;\subseteq \Ev_i\times \Ev_i$ to be the relation such that
$e\rightarrow_i e'$ if $e<_i e'$ and there is no $e''$ such that $e<_i
e''<_i e'$. As a consequence, we have that $\leq_i \;=\; \rightarrow_i^*$,
i.e., $\leq_i$ is the reflexive and transitive closure of $\rightarrow_i$.

A \emph{distributed life-cycle} is a family
$\lambda=\{\lambda_i\}_{i\in\Id}$ of local life-cycles such that $\leq =
(\bigcup_{i\in\Id}\leq_i)^*$ defines a partial order of \textit{global
causality} on the set of all events $\Ev=\bigcup_{i\in\Id}\Ev_i$. 

\emph{Communication} is modeled by event sharing, and thus for some event
$e$ we may have $e\in \Ev_i\cap \Ev_j$, with $i\neq j$. In that case,
requiring $\leq$ to be a partial order amounts to requiring that the local
orders are globally compatible, thus excluding the existence of another
$e'\in \Ev_i\cap \Ev_j$ such that $e<_i e'$ but $e'<_j e$. We denote by
$\Ids(e)$ the set $\{i\in\Id\mid e\in\Ev_i\}$, for each $e\in\Ev$.

We can check the progress of an agent by collecting all the local events
that have occurred up to a given point. This yields the notion of the
\emph{local state} of agent $i$, which is a finite set $\xi_i \subseteq
\Ev_i$ down-closed for local causality, i.e., if $e \leq_i e'$ and $e' \in
\xi_i$ then also $e \in \xi_i$. The set $\Xi_i$ of all local states of an
agent $i$ is totally ordered by inclusion and has $\emptyset$ as the
minimal element.

Each non-empty local state $\xi_i$ is reached, by the occurrence of an
event that we call $\last(\xi_i)$, from the local state $\xi_i \setminus
\{\last(\xi_i)\}$. The local states of each agent are totally ordered, as
a consequence of the total order on local events. Since they are discrete
and well-founded, we can enumerate them as follows: $\emptyset$ is the
$0^{\tx{th}}$ state; $\{e\}$, where $e$ is the minimum of
$\tuple{\Ev_i,\leq_i}$, is the $1^{\tx{st}}$ state; and if $\xi_i$ is the
$k^{\tx{th}}$ state of agent $i$ and $\last(\xi_i)\rightarrow_i e$, then
$\xi_i\cup\{e\}$ is agent $i$'s $(k+1)^{\tx{th}}$ state.

We will denote by $\xi_i^k$ the $k^{\tx{th}}$ state of agent $i$, so
$\xi_i^0=\emptyset$ is the initial state and $\xi_i^k$ is the state
reached from the initial state after the occurrence of the first $k$
events. In fact, $\xi_i^k$ is the only state of agent $i$ that contains
$k$ elements, i.e., where $|\xi_i^k|=k$. Given $e\in \Ev_i$,
$e\!\downarrow_i=\{e'\in \Ev_i\,|\,e'\leq_i e\}$ is always a local state.
Moreover, if $\xi_i$ is non-empty, then $\last(\xi_i)\!\downarrow_i=\xi_i$.

We can also define the notion of a \emph{global state}: a finite set
$\xi\subseteq \Ev$ closed for global causality, i.e.~if $e\leq e'$ and
$e'\in\xi$, then also $e\in\xi$. The set $\Xi$ of all global states
constitutes a lattice under inclusion and has $\emptyset$ as the minimal
element. Clearly, every global state $\xi$ includes the local state
$\xi|_i=\xi\cap \Ev_i$ of each agent $i$.  Given $e\in \Ev$,
$e\!\!\downarrow=\{e'\in \Ev \mid e'\leq e\}$ is always a global state.

\begin{figure}
\begin{center}
\begin {tikzpicture}[
-latex,->,>=stealth',shorten >=1pt
]
\node (i) at (0,2)    	{$i$};
\node (j) at (0,1)		{$j$};
\node (k) at (0,0)      {$k$};
\node (e1) at (1,2)      {$e_1$};
\node (e2) at (1,1)      {$e_2$};
\node (e3) at (1,0)      {$e_3$};
\node (e4i) at (2.5,2)      {$e_4$};
\node (e4j) at (2.5,1)      {$e_4$};
\node (e4k) at (2.5,0)      {$e_4$};
\node (e5) at (4,2)      {$e_5$};
\node (e6) at (4,0)      {$e_6$};
\node (e7j) at (5.5,1)      {$e_7$};
\node (e7k) at (5.5,0,0)      {$e_7$};
\node (e8i) at (7,2)      {$e_8$};
\node (e8j) at (7,1)      {$e_8$};
\node (e9) at (7,0)      {$e_9$};
\node(e10i) at (8.5,2)     {$\dots$};
\node(e10j) at (8.5,1)     {$\dots$};
\node(e10k) at (8.5,0)     {$\dots$};
\path (e1) edge  (e4i);
\path (e4i) edge  (e5);
\path (e5) edge  (e8i);
\path (e8i) edge  (e10i);
\path (e2) edge  (e4j);
\path (e4j) edge  (e7j);
\path (e7j) edge  (e8j);
\path (e8j) edge  (e10j);
\path (e3) edge  (e4k);
\path (e4k) edge  (e6);
\path (e6) edge  (e7k);
\path (e7k) edge  (e9);
\path (e9) edge  (e10k);
\path (e4i) edge[dotted,-] (e4j);
\path (e4j) edge[dotted,-] (e4k);
\path (e7j) edge[dotted,-] (e7k);
\path (e8i) edge[dotted,-] (e8j);
\end{tikzpicture}
\end{center}
\caption{A distributed life-cycle for agents $i$, $j$ and $k$.}\label{fig:dist-lc}
\end{figure}
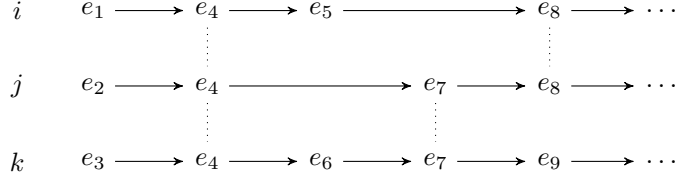

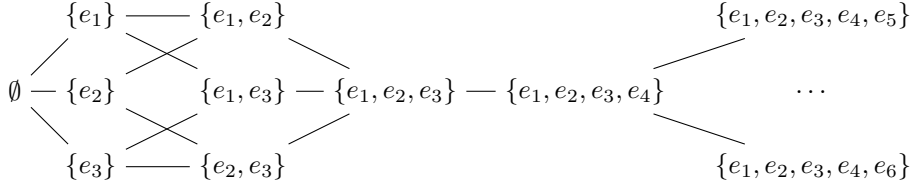
\begin{figure}
\begin{center}
\begin {tikzpicture}[
-latex,-
]
\node (1) at (0,1)    	{$\emptyset$};
\node (2) at (1,2)		{$\{e_1\}$};
\node (3) at (1,1)      {$\{e_2\}$};
\node (4) at (1,0)      {$\{e_3\}$};
\node (5) at (3,2)      {$\{e_1,e_2\}$};
\node (6) at (3,1)      {$\{e_1,e_3\}$};
\node (7) at (3,0)      {$\{e_2,e_3\}$};
\node (8) at (5,1)      {$\{e_1,e_2,e_3\}$};
\node (9) at (7.5,1)      {$\{e_1,e_2,e_3,e_4\}$};
\node (10) at (10.5,2)      {$\{e_1,e_2,e_3,e_4,e_5\}$};
\node (11) at (10.5,0)      {$\{e_1,e_2,e_3,e_4,e_6\}$};
\node (12) at (10.5,1)      {$\dots$};
\path (1) edge  (2);
\path (1) edge  (3);
\path (1) edge  (4);
\path (2) edge  (5);
\path (2) edge  (6);
\path (3) edge  (5);
\path (3) edge  (7);
\path (4) edge  (6);
\path (4) edge  (7);
\path (5) edge  (8);
\path (6) edge  (8);
\path (7) edge  (8);
\path (8) edge  (9);
\path (9) edge  (10);
\path (9) edge  (11);
\end{tikzpicture}
\end{center}
\caption{The lattice of global states.}\label{fig:global-lattice}
\end{figure}

Figure~\ref{fig:dist-lc} depicts a distributed life-cycle where each 
row comprises the local life-cycle of one agent. In particular, $\Ev_i=\{e_1,e_4,e_5,e_8,\dots\}$ and $\to_i$ corresponds to
the arrows in $i$'s row. We can think of the occurrence of event $e_1$ as leading agent $i$ from its initial state $\emptyset$ to the state
$\{e_1\}$, and the the occurrence of event $e_4$ as leading to state $\{e_1,e_4\}$, and so on. Shared events at communication points are highlighted by the dotted vertical lines. Note that the numbers annotating the events are there only for convenience since, in general, no global total order on events is imposed. Figure~\ref{fig:global-lattice} shows that corresponding lattice of global states.

An \emph{interpretation structure} $\mu=\tuple{\lambda,\val}$ consists of
a distributed life-cycle $\lambda$ and a family $\val=\{\val_i\}_{i\in
\Id}$ of local \emph{labeling functions}, where, for each $i\in\Id$, $\val
_i:\Xi_i\to \wp(\Prop_i)$ associates a set of local state propositions to
each local state. We denote the tuple $\tuple{\lambda_i,\val_i}$ also by
$\mu_i$. 

We can the define a \emph{global satisfaction relation} as follows. Given
a global interpretation structure $\mu$ and a global state $\xi$ then
\begin{itemize}

\item $\mu,\xi\sat \neg\alpha$ if $\mu,\xi\not\sat\alpha$;
\item $\mu,\xi\sat \alpha\limp\beta$ if $\mu,\xi\not\sat\alpha$ or $\mu,\xi,\sat\beta$;
\item $\mu,\xi\sat \gf{i}{\varphi}$ if $\mu_i,\xi|_i\sat_i \varphi$.
\end{itemize}
The \emph{local satisfaction relations} at local states are defined by
\begin{itemize}
\item $\mu_i,\xi_i\sat_i p$ if $p\in\val_i(\xi_i)$;
\item $\mu_i,\xi_i\sat_i \lneg\varphi$ if 
      $\mu_i,\xi_i\not\sat_i \varphi$;
\item $\mu_i,\xi_i\sat_i \varphi\limp\psi$ if 
      $\mu_i,\xi_i\not\sat_i \varphi$ or $\mu_i,\xi_i\sat_i \psi$;
\item $\mu_i,\xi_i\sat_i \nxt\varphi$ if there is $e\! \in \!\Ev_i\setminus \xi_i$ such that $\xi_i\cup\{e\}\!\in\!\Xi_i$ and $\mu_i,\xi_i\cup\{e\}\sat_i \varphi$;
\item $\mu_i,\xi_i\sat_i \always\varphi$ if
      $\mu_i,\xi'_i\sat_i \varphi$, for every $\xi'_i\in\Xi_i$ such that 
      $\xi_i\subseteq \xi'_i$;
\item $\mu_i,\xi_i\sat_i \comm{j}{\varphi}$ if 
      $\xi_i\neq\emptyset$, $\last(\xi_i)\in\Ev_j$ and 
      $\mu_j,\last(\xi_i)\!\!\downarrow_j\sat_j {\varphi}$. 
\end{itemize}

We say that $\mu$ \emph{(globally) satisfies} $\alpha$, or that $\mu$ is a \textit{model} of $\alpha$, written
$\mu\sat\alpha$, whenever $\mu,\emptyset\sat\alpha$. As expected, $\alpha$
is said to be \emph{satisfiable} whenever there is $\mu$ such that
$\mu\sat\alpha$. We denote by $\Mod(\alpha)$ the set of all models of $\alpha$. 
We define  similar notion  for the local languages. We say that $\mu_i$ \emph{(locally) satisfies} $\varphi$, written $\mu_i\sat_i\varphi$ if $\mu_i,\emptyset\sat_i\varphi$.

The following result will be useful in the future. It basically captures the traditional fixed-point characterization of the $\always$ temporal  operator:
$$
\always \varphi\leqv (\varphi\lconj \nxt\always \varphi).
$$

\begin{lemma}\label{lemma:always-fp}
Let $\mu_i$ be a local interpretation structure and $\xi_i\in\Xi_i$ any of its local states. Then
\begin{enumerate}
\item $\mu_i,\xi_i\sat_i \always \varphi$ iff $\mu_i,\xi_i\sat_i \varphi$ and $\mu_i,\xi_i\cup\{e\}\sat_i \always\varphi$, provided that $\xi_i\cup\{e\}\in\Xi_i$.
\end{enumerate}
\end{lemma}

\section{Distributed B\"uchi Automata}\label{sec:dba}

In this section we present B\"uchi automota for \DTL. We start by presenting the traditional notion of nondeterministic B\"uchi automaton and generalized nondeterministic B\"uchi automaton. We use these notions to capture the local behaviour of the agents, given that each agent is essentially linear. In this case, we follow very closely the ideas presented in \cite{bai:kat:08}. Then, we propose a novel notion of \textit{distributed B\"uchi automaton} to capture the distributed nature of DTL.

A \textit{nondeterministic B\"uchi automation} (NBA) is a tuple $\cA=\tuple{Q,\Sigma,\delta,Q_0,F}$ where:
\begin{itemize}
\item $Q$ is a nonempty finite set of \textit{states};

\item $\Sigma$ is a finite set \textit{alphabet symbols} such that $Q\cap\Sigma=\emptyset$;

\item $\delta:Q\times\Sigma \to 2^Q$ is the \textit{transition function};

\item $Q_0\subseteq Q$ is a set of \textit{initial states};

\item $F\subseteq Q$ is a set of \textit{acceptance states} (also called \textit{final states}).
\end{itemize}
When $q'\in\delta(q,a)$, we may write $q\transact{a}q_{k'}$ instead. Let $\Sigma^\omega$ denote the set of all infinite words over $\Sigma$.
A \textit{run} for $w=a_0a_1a_2\dots\in\Sigma^\omega$ in $\cA$ is an infinite sequence $q_0q_1q_2\dots$ of states in $\cA$ such that 
$q_0\in Q_0$ and $q_k\transact{a_k}q_{k+1}$, for $k\in\nats$:
$$
q_0\transact{a_0} q_1\transact{a_1} q_2\transact{a_2}\dots
$$
A run $q_0q_1q_2\dots$ is \textit{accepting} if $q_k\in F$ for infinitely many indices $k\in\nats$. The \textit{accepted language} of $\cA$ is
$$
L(\cA)=\{w\in \Sigma^\omega\mid \textrm{there exists an accepting run for }w\text{ in }\cA\}.
$$

A \textit{generalized nondeterministic B\"uchi automaton} (GNBA) is a tuple $\cG=\tuple{Q,\Sigma,\delta,Q_0,\cF}$ where $Q$, $\Sigma$, $\delta$ and $Q_0$ are defined just as for NBA and $\cF$ is a (possibly empty) subset of $2^Q$. The elements of $\cF$ are called \textit{acceptance sets}. A \textit{ run} for $w=a_0a_1a_2\dots\in\Sigma^\omega$ in $\cG$ is defined as in the case of an NBA. A run $q_0q_1q_2\dots$ is \textit{accepting} if for each acceptance set 
$F\in \cF$ there are infinitely many indices $k\in \nats$ such that $q_k\in F$. The accepted language for a GNBA is defined just as for the case of an NBA.

The classes of NBA's and GNBA's are \textit{equivalent} in the sense that they accept exactly the same languages. 
Every NBA is a particular case of a GNBA. Furthermore, for each GNBA $\cG$ there exists an NBA $\cA_{\cG}$ such that 
$L(\cA_{\cG})=L(\cG)$. Details of this equivalence can be found in~\cite{bai:kat:08}.

In the sequel, we overload the $\,{}\da_i$ notation and use $q\da_i$ to denote the projection of tuple $q$ over component $i$.

Next, we present the novel notion of distributed B\"uchi automata for  $\DTL$. From now on, we assume fixed a distributed signature $\Sigma=\tuple{\Id,\{\Prop\}_{i\in\Id}}$. For each $i\in \Id$, let $\cA_i=\tuple{Q_i,\Sigma_i,\delta_i,Q_{0_i},F_i}$ be an NBA such that for distinct $i,j\in \Id$:
\begin{itemize}
\item $Q_i\cap Q_j=\emptyset$;
\end{itemize}
A \textit{distributed nondeterministic B\"uchi automaton} (DNBA)  based on $\{\cA_i\}_{i\in\Id}$ is a tuple
$$
\cD=\tuple{Q,\Sigma,\delta,Q_0,\cF}
$$
such that:
\begin{itemize}
\item $Q=\bigotimes_{i\in\Id}Q_i$;

\item $\Sigma=\{a\subseteq \uplus_{i\in\Id}\Sigma_i\mid a\neq\emptyset\textrm{ and }|a\cap \Sigma_i|\leq 1\textrm{, for }i\in \Id\}$;

\item $\delta:Q\times\Sigma\to 2^Q$ is such that $\delta(q,a)$ is the set of all states $q'$ satisfying:
	\begin{itemize}
	\item if $a\cap \Sigma_i=\emptyset$ then $q'\da_i=q\da_i$;
	\item if $a\cap \Sigma_i\neq\emptyset$ then $q'\da_i\in\delta_i(q\da_i,a\cap \Sigma_i)$;
	\end{itemize}

\item $Q_0=\bigotimes_{i\in\Id}Q_{0_i}$;

\item $\cF=\{\cF_i\subseteq Q\mid\textrm{ for }i\in\Id\}$ such that $\cF_i=\{q\in Q\mid q\da_i\in F_i\}$;
\end{itemize}

The states of the DNBA are tuples of states from the local automata, one for each agent.
Each symbol of the distributed alphabet is a nonempty set of symbols of the local automata with the proviso that in each global symbol there is at most one symbol from each agent. 
The transition from one state to the next at the global level is guided by the local behaviour of each component. If, for a particular global symbol $a$, agent $i$ is not involved, that is, if $a\cap\Sigma_i=\emptyset$ then for this transition the agent's local state will not change. If, on the other hand, the agent is involved in $a$, that is, if $a\cap\Sigma_i\neq\emptyset$ then the agent's local state will change according to its local behaviour, which is dictated by $\delta_i$. Note that, in this case, we are abusing notation. If $a\cap\Sigma_i\neq\emptyset$ then $a\cap\Sigma_i$ is a set, a singleton $\{a'\}$ with $a'\in\Sigma_i$, but nevertheless, a set. Hence, when we write $\delta_i(q\da_i,a\cap \Sigma_i)$ we  obviously mean $\delta_i(q\da_i,a')$.

The language accepted by the distributed automaton will be as expected. It will accepted all the local words of the local automata. However, we need one additional proviso: we only consider \textit{fair} words. A global word $a_0a_1a_2$ is \textit{fair} if, for every $i\in\Id$, $a_k\cap \Sigma_i\neq\emptyset$, for infinitely many indices $k\in\nats$. We need to ensure that a global accepting run is locally accepting for each agent. So, we cannot simply promote a state $q$ to accepting because one of its components is accepting in the local automaton. This would allow for the acceptance of other words from other local automata. A \textit{global run} for a fair word $w$ in $\cD$ is a sequence of states $q_0q_1q_2\dots$ such that $q_k\transact{a_k}q_{k+1}$, just as for the local case. A global run $a_0a_1a_2\dots$ is \textit{accepting} if, for each $i\in\Id$, $q_k\in \cF_i$, for infinitely many indices $k\in\nats$.

Let $w=a_0a_1a_2\dots \in \Sigma^\omega$ be a fair global word. Then, we denote by $w\da_i$ the local word obtained from $w$ as follows:
\begin{itemize}
\item first, consider the projection $w'=(a_0\cap \Sigma_i) (a_1\cap \Sigma_i) (a_2\cap \Sigma_i)\dots$ over the alphabet $\Sigma_i$;
\item then, let $w\da_i$ be the local word obtained from $w'$ by removing all the empty sets and replacing each nonempty set $\{a\}$ by its element $a$.
\end{itemize}
Recall that $|a_k\cap\Sigma_i|\leq1$ hence $a_k\cap\Sigma_i$ is either a singleton or the empty set.

Similarly, let  $\tau=q_0q_1q_2\dots$ be  a global run for $w=a_0a_1a_2\dots$. Then, we denote by $\tau\da_{w,i}$ the local run for $w\da_i$ obtained from $\tau$ as follows:
\begin{itemize}
\item first, consider the projection $\tau'=q_0\da_iq_1\da_iq_2\da_i\dots$ over the states of $\cA_i$;
\item then, let $\tau\da_{w,i}$ be the local run obtained from $\tau'$ by removing $q_{k+1}$ if $(a_k\cap \Sigma_i)=\emptyset$, for $k\in\nats$.
\end{itemize}
In this case, we project each global state on its local component for agent $i$ and then remove all the states resulting from transitions where agent $i$ was not involved. The follow lemma proves that $w\da_i$ is indeed a word in $\Sigma_i^\omega$ and that $\tau\da_{w,i}$ is a local run for $w\da_i$ in $\cA_i$.

\begin{lemma}\label{dnba:localvsglobal}
Let $\cD$ be a DNBA based on $\{\cA_i\}_{i\in\Id}$. If $\tau=q_0q_1q_2\dots$ is a global run for a fair global word $w=a_0a_1a_2\dots \in \Sigma^\omega$ then, for each $i\in\Id$:
\begin{enumerate}
\item $w\da_i\in\Sigma_i^\omega$;
\item $\tau\da_{w,i}$ is a local run for $w\da_i$ in $\cA_i$.
\end{enumerate}
Furthermore, $\tau$ is accepting if and only if $\tau\da_{w,i}$ is accepting, for every $i\in\Id$. 
\end{lemma}
\proof{
1.~Straightforward from the definition of $w\da_i$ and the fact that $w$ is fair.\\
2.~We briefly sketch the intuition behind this result. Consider the situation:
$$
\dots\tuple{\dots,q_k,\dots}\transact{a_k}\tuple{\dots,q_{k+1},\dots}\transact{a_{k+1}}\tuple{\dots,q_{k+2},\dots}\dots
$$
where $a_k\cap\Sigma_i=\emptyset$ and $a_{k+1}\cap\Sigma_i\neq\emptyset$. Then, in $\tau\da_{w,i}$, we get the following situation:
$$
\dots q_k \transact{a_{k+1}\cap\Sigma_i}q_{k+2}\dots
$$
where $q_{k+1}\da_i$ was deleted. By definition of $\delta$, 
$$
q_{k+1}\da_i=q_{k}\da_i \textrm {and }q_{k+2}\da_i\in\delta_i(q_{k+1}\da_i,a_{k+1}\cap\Sigma_i).
$$
It is not very difficult to conclude that this leads to a run in $\cA_i$. Furthermore, the fact that  $\tau$ is accepting if and only if $\tau\da_{w,i}$ is accepting, for every $i\in\Id$, is an immediate consequence of the definition of global acceptance.
}

\begin{example}
Consider the NBA's $\cA_1$ and $\cA_2$ depicted in Figure~\ref{example:nbas}, with $\Sigma_1=\{0,1\}$ and $\Sigma_2=\{a,b\}$.  $\cA_1$ accepts all the infinite words over $\{0,1\}$ with infinitely many 0's, and $\cA_2$ accepts all the infinite words over $\{a,b\}$ with finitely many $a$'s.
\begin{figure}[ht]
\begin{center}
\scalebox{0.8}{
\begin {tikzpicture}[-latex,->,>=stealth',shorten >=1pt,auto,semithick,initial text=$ $]
\node[state,initial] (q0)   at    (0,2)      {$q_0$};
\node[state,accepting] (q1)   at    (2,2)    {$q_1$};
\node[state,initial] (p0)	  at	  (7,2)    {$p_0$};
\node[state,accepting] (p1)    at    (9,2)    {$p_1$};
\draw (q0) edge[bend left] node[above] {\small$0$} (q1)
      (q0) edge[loop above] node[above] {\small$1$} (q0)
	  (q1) edge[loop above] node[above] {\small$0$} (q1)
	  (q1) edge[bend left] node[below] {\small$1$} (q0)
	  (p0) edge[loop above] node[above] {\small$a,b$} (p0)
	  (p0) edge node[above]{\small$b$} (p1)
	  (p1) edge[loop above] node[above]{\small$b$} (p1);
\end{tikzpicture}
}
\end{center}
\caption{NBA's $\cA_1$ (on the left) and $\cA_2$ (on the right).}\label{example:nbas}
\end{figure}
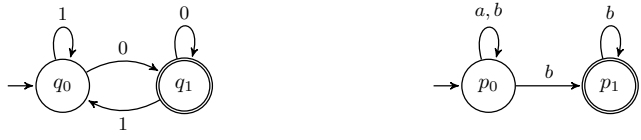

Now let us consider the DNBA $\cD$ based on $\{\cA_1,\cA_2\}$. The alphabet is composed of sets with one symbol from one or from the two agents:
$$\Sigma=\{\{0\},\{1\},\{a\},\{b\},\{0,a\},\{0,b\},\{1,a\},\{1,b\}\}.$$ The set of states is 
$$
Q=\{\tuple{q_0,p_0},\tuple{q_0,p_1},\tuple{q_1,p_0},\tuple{q_1,p_1}\}.
$$
Of these, only $\tuple{q_0,p_0}$ is initial, and 
$$F_1=\{\tuple{q_1,p_0},\tuple{q_1,p_1}\}\textrm{ and }F_2=\{\tuple{q_0,p_1},\tuple{q_1,p_1}\}.$$
The transition function $\delta$ is depicted is Figure~\ref{example:dnba}.
\begin{figure}[ht]
\begin{center}
\scalebox{0.8}{
\begin {tikzpicture}[-latex,->,>=stealth',shorten >=1pt,auto,semithick,initial text=$ $]
\node[state,initial] (q00)   at    (0,4)      {$\tuple{q_0,p_0}$};
\node[state] (q01)   at    (4,4)    {$\tuple{q_0,p_1}$};
\node[state] (q10)	  at	  (0,0)    {$\tuple{q_1,p_0}$};
\node[state] (q11)    at    (4,0)    {$\tuple{q_1,p_1}$};
\draw (q00) edge node[above] {\small$\{b\},\{1,b\}$} (q01)
      (q00) edge[loop above] node[above] {\small$\begin{array}{c}\{1\},\{a\},\{b\}\\\{1,a\},\{1,b\}\end{array}$} (q00)
      (q00) edge node[right=2pt,near start] {\small$\{0,b\}$} (q11)
      (q00) edge[bend left] node[right=-4pt,near end] {\small$\begin{array}{c}\{0\}\\\{0,a\}\\\{0,b\}\end{array}$} (q10)
	  (q01) edge[loop above] node[above] {\small$\{1\},\{b\},\{1,b\}$} (q01)
	  (q01) edge[bend left] node[right] {\small$\begin{array}{c}\{0\}\\\{0,b\}\end{array}$} (q11)
	  (q10) edge[loop below] node[below]{\small$\begin{array}{c}\{0\},\{a\},\{b\}\\\{0,a\},\{0,b\}\end{array}$} (q10)
	  (q10) edge[bend left] node[left=-2pt] {\small$\begin{array}{c}\{1\}\\\{1,a\}\\\{1,b\}\end{array}$} (q00)
	  (q10) edge node[above]{\small$\{b\},\{0,b\}$} (q11)
	  (q11) edge[bend left] node[left=-6pt] {\small$\begin{array}{c}\{1\}\\\{1,b\}\end{array}$} (q01)
	  (q11) edge[loop below] node[below]{\small$\{0\},\{b\},\{0,b\}$} (q11);
\end{tikzpicture}
}
\end{center}
\caption{DNBA $\cD$ based on $\{\cA_1,\cA_2\}$.}\label{example:dnba}
\end{figure}
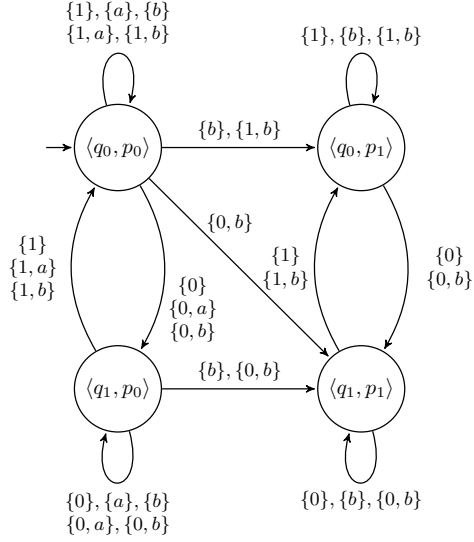

It is not very difficult to observe that words with an infinite number of 0's and a finite number of $a$'s are accepted. For instance, the word
$$
\{1\}\{a\}\{1,b\}\{0,b\}\{0\}\{b\}\{0,b\}\{0\}\{b\}\dots
$$
is accepted given that state $\tuple{q_1,p_1}$ is visited infinitely often, that is, a final state form $\cA_1$ and a final state from $\cA_2$ are visited infinitely often. However, if this was the only requirement for acceptance, then the word
$$
\{1\}\{a\}\{1,b\}\{0,b\}\{0\}\{0\}\{0\}\{0\}\dots
$$
would also be accepted given that state $\tuple{q_1,p_1}$, in this case, is also visited infinitely often. We don't want this word to be accepted because its projection on $\cA_2$ yields the finite word $aba$ which is not part of the language of $\cA_2$. But, this global word is not a fair word hence it will not be accepted by $\cD$.
\end{example}

Our goal now is to define a DNBA $\cD_\alpha$ for a given formula 
$\alpha\in\LL$ that accepts all the models of $\alpha$ and only those. To this end, we start by defining some auxiliary notions.

From now on, assume fixed a global formula $\alpha$. A set $B\subseteq\closure(\alpha)$ is said to be \textit{consistent} with respect to propositional logic if:
\begin{itemize}

\item $\gamma_1\limp\gamma_2\in B$ if and only if $\lneg\gamma_1\in B$ or $\gamma_2\in B$, for $\gamma_1\limp\gamma_2\in \closure(\alpha)$;

\item if $\gamma_1\in B$ then $\lneg\gamma_1\notin B$;

\item if $\top\in\closure(\alpha)$ then $\top\in B$.

\end{itemize}
Herein, $\gamma_1,\gamma_2$ denote either local or  global formulas.

A set $B\subseteq\closure(\alpha)$ is said to be \textit{locally consistent} with respect to the temporal operator $\always$ if:
\begin{itemize}
\item if $\always\varphi_1\in B$ then $\varphi_1\in B$, for every $\always\varphi_1\in\closure(\alpha)$.

\end{itemize}

A set $B\subseteq\closure(\alpha)$ is said to be \textit{$i$-consistent} with respect to global formulas if
\begin{itemize}
\item $\gf{i}{\varphi}\in B$ iff $\varphi\in B$, for every $\gf{i}{\varphi}\in\subf(\alpha)$.
\end{itemize}

A set $B\subseteq\closure(\alpha)$ is said to be \textit{maximal} if for all $\gamma\in\closure(\alpha)$:
\begin{itemize}
\item if $\gamma\notin B$ then $\lneg \gamma\in B$.
\end{itemize}

A set $B\subseteq\closure(\alpha)$ is $i$-\textit{elementary} if it is consistent with respect to propositional logic, maximal and locally consistent with respect to the temporal operator $\always$ and $i$-consistent with respect to global formulas. Elementary sets try to capture all the properties that can be asserted locally. When $i$ is clear from context, we may write elementary instead of $i$-elementary.
 
Recall formula $\gf{i}{\nxt(p\limp\comm{j}{q})}\limp\gf{j}{\nxt q}$. The set
$$
\begin{array}{rcl}
B_1&=&\{\gf{i}{\nxt(p\limp\comm{j}{q})}\limp\gf{j}{\nxt q},\gf{i}{\nxt(p\limp\comm{j}{q})},\gf{j}{\nxt q},\\[2mm]
 &&\;\;\;\nxt(p\limp\comm{j}{q}),\nxt q,\lneg\nxt p,\lneg\comm{j}{q},q\}
\end{array}
$$
is an example of an elementary sets. However, for instance, sets 
$$
B_2=\{\gf{i}{\nxt(p\limp\comm{j}{q})}\limp\gf{j}{\nxt q},\gf{i}{\nxt(p\limp\comm{j}{q})},\lneg\gf{j}{\nxt q},\dots\}
$$
and
$$
B_3=\{\gf{i}{\nxt(p\limp\comm{j}{q})}\limp\gf{j}{\nxt q},\lneg\gf{i}{\nxt(p\limp\comm{j}{q})},\gf{j}{\nxt q},\lneg\nxt q,\dots\}
$$
are not elementary. Set $B_2$ is not consistent with propositional logic and set $B_3$ is not $j$-consistent with global formulas.

We have all we need to define the envisaged DNBA. We start by defining the local GNBA's $\cG_i$ for each agent $i\in\Id$. The construction is similar  the one presented in \cite{bai:kat:08}. From these, we can then obtain equivalent NBAs $\cA_i$ that will be used to define the DNBA. Each GNBA  $\cG_i=\tuple{Q_i,\Val_i,\delta_i,Q_{0_i},\cF_i}$ is as follows:
\begin{itemize}
\item $Q_i=\{B\da_i\mid B\subseteq \closure(\alpha)\textrm{ and } B\textrm{ is $i$-elementary}\}$;

\item $Q_{0_i}=\{B\in Q_i\mid\alpha\in B\textrm { and }\comm{j}{\varphi}\notin B\textrm{, for }\comm{j}{\varphi}\in\closure(\alpha)\}$;

\item $\cF_i=\{F_{\always\varphi}\mid \always\varphi\in \closure(\alpha)\}$ 
where 
\begin{itemize}
\item $F_{\always\varphi}=\{B\in Q_i\mid \always\varphi\in B \textrm{ or } \varphi\notin B\}$;
\end{itemize}

\item $\delta_i:Q_i\times \Val_i\to 2^{Q_i}$ is such that:
	\begin{itemize}
	\item if $v\neq B\cap \lit_i$ then $\delta_i(B,v)=\emptyset$;
	\item if $v= B\cap \lit_i$ then $\delta_i(B,v)$ is the set of all elementary sets $B'$ such that:
		\begin{enumerate}
		\item $\nxt \varphi\in B$ iff $\psi\in B'$, for every $\nxt \varphi\in\closure(\alpha)$;
		\item $\always \varphi\in B$ iff $\varphi\in B$ and $\always\varphi\in B'$, for every $\always \varphi\in\closure(\alpha)$.
		\end{enumerate}
	\end{itemize}
\end{itemize}
Recall that $\Val_i$ is the set of all $i$-valuations and that a valuation is a set of literals such that, for each propositional symbol either the symbol is in the valuation or its negation is. This is the alphabet of the automaton. The states of the automaton are all the elementary sets (restricted to the relevant formulas, that is, all the global formulas and all the local formulas for the agent at hand). Each state contains all the formulas that are intended to hold at that point. In particular, initial states characterize the initial set-up conditions. We want $\alpha$ to hold initially and, as imposed by the  semantics of DTL, there can only by synchronizations after the first event occurs. Hence, there can be no communication formulas in any initial state. Regarding the transition function, given an alphabet symbol $v$ and state $B$, the transition will only be enabled if the valuation $v$ \textit{agrees} with the information in $B$, i.e., if the state propositions in $v$ are also present in $B$, meaning that they should be true, and  the negation of state propositions in $v$ are also present in $B$, meaning that they should be false. Additionaly,
conditions (1) and (2)  reflect the semantics of temporal operators. In particular, condition (2) is based on the fixed-point semantics of the $\always$ operator. The final states are defined in order to capture the temporal semantics of the $\always$ operator. They are basically used to exclude runs where, from a certain point on, a formula $\varphi$ is always true (that is, it present in all the states) but $\always\varphi$ is not (that is, it is not present in the states of the run), for instance, as follows:
$$
B_0\transact{v_0}\dots\transact{v_{k-1}}\{\dots,\varphi,\dots\}\transact{v_{k}}\{\dots,\varphi,\dots\}\transact{v_{k+1}}\dots
$$
In this run, $\varphi$ is present in all the states starting from $k$. Then, this means that $\always\varphi$ is true from that point on. In order for the run to be accepting, $F_{\always\varphi}$ must be visited infinitely often. This means that after $k$ and as $\varphi\in B_n$ for $n\geq k$ then $\always \varphi$ must be in infinitely of these states. Then, by condition (2) of the transition function, $\always \varphi$ must be in all of them, as intended.

For each $i\in\Id$, let $\cA_i$ be an NBA equivalent to the GNBA $\cG_i$.
Then $\cD_\alpha=\tuple{Q,\Sigma,\delta,Q_0,\cF}$ is the DNBA based on $\{\cA_i\}_{i\in\Id}$ satisfying the following conditions, for every $q,q'\in Q$, $a\in \Sigma$ and $i,j\in\Id$:
\begin{description}[leftmargin=!,labelwidth=\widthof{\bfseries (SC3)},align=right]

\item[\namedlabel{dnba:lc}{(LC)}] if  $\comm{j}{\varphi}\in q\da_i$ then $\varphi\in q\da_j$;

\item[\namedlabel{dnba:sync1}{(SC1)}]  if $q'\in\delta(q,a)$ and  $\comm{j}{\varphi}\in q'\da_i$ and $a\cap\Val_i\neq\emptyset$ then $a\cap\Val_j\neq\emptyset$;

\item[\namedlabel{dnba:sync2}{(SC2)}] if $a\cap\Val_i\neq\emptyset$ and $a\cap\Val_j\neq\emptyset$ and $\varphi\in q'\da_j$ then $\comm{j}{\varphi}\in q'\da_i$.
\end{description}
Condition~\ref{dnba:lc} states that in a global state, if for agent $i$ it holds $\comm{j}{\varphi}$ then it must be the case that  $\varphi$ holds for agent $j$, as intended for the semantics of the communication primitive. Condition~\ref{dnba:sync1} states that in every state $q'$ reached by a transition where $i$ was an active participant (expressed by $a\cap\Val_i\neq\emptyset$), if $\comm{j}{\varphi}$ holds for $i$ then $i$ and $j$ must have just synchronized and so $j$ must also have been an active participant in $a$ (expressed by $a\cap\Val_j\neq\emptyset$). Finally, condition~\ref{dnba:sync1} states that if $i$ and $j$ were both active in the last transition and if $\psi$ holds for agent $j$ then, in the event that $i$ wants to communicate with $j$, it will be able to infer that $\psi$ holds for $j$, that is, $\comm{j}{\psi}$ holds for $i$.

We now proceed to show the correctness of this construction. We aim at proving that any word accepted by the automaton is captured by a DTL model of $\alpha$ and that any DTL model of $\alpha$ is represented by words accepted by the automaton.

Let  $w=a_0a_1a_2\dots\dots\in L(\cD_\alpha)$. Then, there is an accepting run $\tau=q_0q_1q_2\dots$ in $\cD_\alpha$. We denote by $\mu^{\tau}=\tuple{\lambda,\vartheta}$ the  interpretation structure induced by $\tau$ (and consequently by $w$), defined as follows:
\begin{itemize}
\item $\Ev=\{e_k\mid k\geq1\}$;
\item $\Ev_i=\{e_k\in \Ev\mid a_{k-1}\cap \Val_i\neq\emptyset\}$;

\item $\lambda_i=\tuple{\Ev_i\leq_i}$ is the local life-cycle such that $e_{k_1}\leq_i e_{k_2}$ if $k_1\leq k_2$;
\item $\lambda=\{\lambda_i\}_{i\in\Id}$ is the corresponding distributed life-cycle;
\item $\vartheta_i:\Xi_i\to2^{\Prop_i}$ is such that, for every $p\in\Prop_i$ and $e_k\in\Ev_i$:
	\begin{itemize}
	\item $\vartheta_i(\emptyset)=\begin{cases}
										1&\textrm{if }p\in q_0\da_i\\
										0&\textrm{if }\lneg p \in q_0\da_i
									 \end{cases}$
	\item $\vartheta_i(\xi\cup\{e_k\})=\begin{cases}
										1&\textrm{if }p\in q_k\da_i\\
										0&\textrm{if }\lneg p \in q_k\da_i
									 \end{cases}$, for $\xi\in\Xi_i$;
	\end{itemize}
\item $\vartheta=\{\vartheta_i\}_{i\in\Id}$.
\end{itemize}
Note that each $\vartheta_i$ is well defined because $q_k$ is elementary. In the sequel, we can consider the following enumeration of global events:
\begin{itemize}
\item $\xi^0=\emptyset$
\item $\xi^k=\{e_1,\dots,e_k\}$, for $k\geq1$.
\end{itemize}

\begin{theorem}
If $w\in L(\cD_{\alpha})$ then $\mu^{\tau}\in\Mod(\alpha)$, for some accepting run $\tau$ for $w$ in $\cD_\alpha$.
\end{theorem}
\proof{Let $\tau=q_0q_1q_2\dots$.
Our goal is to prove that $\mu^{\tau}\sat\alpha$. We start by establishing a preliminary result for the local level. 
 We prove that  $\mu^{\tau}_i,\xi^k|_i\sat_i\psi$ iff $\psi\in q_k\da_i$, for every $\psi\in \closure(\alpha)\cap\LL_i$. The proof  is done  by induction on the structure of $\psi$, simultaneously for all agents. \\[2mm]
Basis: $\psi\in\Prop_i$. Then, if $k=0$ then $\xi^0=\emptyset$ and $p\in q_0\da_i$ iff $\vartheta_i(\emptyset)=1$ iff $\mu^{\tau}_i,\xi^0\sat_i p$. If $k>0$, then $\xi^k\neq\emptyset$ and $\last(\xi^k)=e_k$. If $e_k\in\Ev_i$ then $\xi^k|_i=\xi'\cup\{e_k\}$, for some local state $\xi'\in\Xi_i$, and $p\in q_k\da_i$ iff
$\vartheta_i(\xi'\cup\{e_k\})=1$ iff $\mu^{\tau}_i,\xi^k|_i\sat_i p$. If $e_k\notin \Ev_k$, and consequently 
$a_{k-1}\cap\Val_i=\emptyset$, then we distinguish two cases: either (i) 
$a_{k_1}\cap\Val_i=\emptyset$, for every $k_1<k$; or 
(ii) there is $k_1<k$ such that $\Val_{k_1}\cap\Val_i\neq\emptyset$. In the first case, this means that $e_{k_1}\notin \Ev_i$, for $k_1<k$ and so $\xi^k|_i=\emptyset$ and, by definition of DNBA, it also follows that $q_0\da_i=q_1\da_i=\dots= q_k\da_i$. The proof then follows as in the case of $k=0$. In case (ii), let $k_1$ be the greatest $k'<k$ such that $a_{k'}\cap\Val_i\neq\emptyset$. Then, $e_{k_1+1}\in\Ev_i$, $\last(\xi^k|_i)=e_{k_1+1}$ and $\xi^k|_i=\xi'\cup\{e_{k_1+1}\}$, for some local state $\xi'\in\Xi_i$. Furthermore, it follows by definition of DNBA that $q_{k_1+1}\da_i=\dots=q_k\da_i$. Hence, $p\in q_{k}\da_i$ iff $p\in q_{k_1+1}\da_i$ iff
$\vartheta_i(\xi'\cup\{e_{k_1+1}\})=1$ iff $\mu^{\tau}_i,\xi'\cup\{e_{k_1+1}\}\sat_i p$ iff $\mu^{\tau}_i,\xi^k|_i\sat_i p$.  \\[2mm]
Induction step: The case of propositional formulas is an immediate consequence of the definition elementary set and we omit the details.\\[2mm]
Assume that $\psi=\nxt\psi_1$. Additionally, assume also that $\nxt\psi_1\in q_k\da_i$. If $a_k\cap\Val_i\neq\emptyset$ then $\psi_1\in q_{k+1}\da_i$, by definition of $\cG_i$, and, by induction hypothesis, $\mu^\tau_i,\xi^{k+1}|_i\sat_i\psi_1$. But  $a_k\cap\Val_i\neq\emptyset$ also implies that $e_{k+1}\in\Ev_i$ and, so, $\xi^{k+1}|_i=(\xi^k\cup\{e_{k+1}\})|_i=\xi^k|_i\cup\{e_{k+1}\}$. Hence, $\mu^\tau_i,\xi^k|_i\cup\{e_{k+1}\}\sat_i\psi_1$ which implies that $\mu^\tau_i,\xi^k|_i\sat_i\nxt\psi_1$. If $a_k\cap\Val_i=\emptyset$ then let $k_1$ be the least index greater than $k$ such that $a_{k_1}\cap\Val_i\neq\emptyset$. Then, by definition of DNBA, $q_k\da_i=\dots=q_{k_1}\da_i$ and so $\nxt\psi_1\in q_{k_1}\da_i$. Consequently, $\psi_1\in q_{k_1+1}\da_i$. By induction hypothesis, $\mu^\tau_i,\xi^{k_1+1}|_i\sat_i\psi_1$. By definition of $\Ev_i$, it follows that $e_{k_1+1}\in \Ev_i$, $\xi^{k_1}|_i=\xi^k|_i$ and thus $\xi^{k_1+1}|_i=(\xi^{k_1}\cup\{e_{k_1+1}\})|_i=\xi^{k_1}|_i\cup\{e_{k_1+1}\}=\xi^{k}|_i\cup\{e_{k_1+1}\}$. Hence, $\mu^\tau_i,\xi^{k}|_i\cup\{e_{k_1+1}\}\sat_i\psi_1$ and thus $\mu^\tau_i,\xi^{k}|_i\sat_i\nxt\psi_1$. Assume now that $\mu^\tau_i,\xi^{k}|_i\sat_i\nxt\psi_1$. Then, there is $e\in\Ev_i\setminus\xi^k|_i$ such that $\xi^k|i\cup\{e\}\in\Xi_i$ and $\mu^\tau_i,\xi^k|i\cup\{e\}\sat_i\psi_1$. Clearly, there is $k_1>k$ such that $e=e_{k_1}$ and $\xi^k|_i\cup\{e\}=\xi^{k_1}|_i$. Using the induction hypothesis, it follows that $\psi_1\in q_{k_1}\da_i$. If $k_1=k+1$ then $a_k\cap\Val_i\neq\emptyset$ and, by definition of $\cG_i$, $\nxt\psi_1\in q_{k}\da_i$. If $k_1>k+1$ then $a_k\cap\Val_i=\dots=a_{k_1-2}\cap\Val_i=\emptyset$ and 
$a_{k_1-1}\cap \Val_i\neq\emptyset$. Hence $q_k\da_i=\dots=q_{k_1-1}\da_i$ and if $\psi_1\in q_{k_1}\da_i$ then $\nxt\psi_1\in q_{k_1-1}\da_i=q_k\da_i$. \\[2mm]
Assume now that $\psi$ is $\always\psi_1$. We start by observing that for any run $B_0B_1B_2\dots$ in $\cG_i$ if $\always\psi_1\in B_k$ then $\always\psi_1\in B_{k'}$ for every $k'\geq k$. This is an immediate consequence of condition (2) in the definition of $\delta_i$ and can easily be established by induction. Assume first that $\always \psi\in  q_k\da_i$. Then, by the previous claim and Lemma~\ref{dnba:localvsglobal}, it follows that $\always\psi_i\in q_{k'}\da_i$ for every $k'\geq k$. As each set $q_{k'}\da_i$ is elementary then it is locally consistent with respect to temporal operator $\always$ and so $\psi_1\in q_{k'}\da_i$, for every $k'\geq k$. Using the induction hypothesis, it follows that $\mu^\tau_i,\xi^{k'}|_i\sat_i\psi_1$, for every $k'\geq k$. And this last condition implies that $\mu^\tau_i,\xi^k|_i\sat_i\always\psi_1$. Assume now that $\mu^\tau_i,\xi^k|_i\sat_i\always\psi_1$. Then, $\mu^\tau_i,\xi'_i\sat_i\psi_1$, for every $\xi'_i\supseteq \xi^k|_i$. A simple inductive argument allows us to conclude that $\mu^\tau_i,\xi^{k'}|_i\sat_i\psi_1$, for every $k'\geq k$. Note that if $e_{k'}\in\Ev_i$ then $\xi^{k'}|_i=\xi'_i
\cup\{e_{k'}\}$ which is in $\Xi_i$ and  satisfies $\xi'_i
\cup\{e_{k'}\}\supseteq \xi^k|_i$. If $e_{k'}\notin\Ev_i$ then $\xi^{k'}|_i=\xi^{k'-1}|_i$ and again $\xi^{k'}|_i\supseteq \xi^k|_i$. Hence, by induction hypothesis, it follows that $\psi_1\in q_{k'}\da_i$, for every $k'\geq k$. As the run is accepting then, some of these states must be in $F_{\always\psi_1}$. Let $q_{k_1}\da_i$, with $k_1\geq k$, be the first of such states. Clearly, it must be the case that $k_1=k$. In fact, if $k_1>k$, given that $\psi\in q_{k_1}\da_i$ then, by condition (2) in the definition of $\delta_i$ this would imply that $\always\psi_1\in q_{k_1-1}\da_i$  forcing $q_{k_1-1}\da_i$ to also be in $F_{\always\psi_1}$ and thus contradicting the fact the $k_1$ was the first final state after $k$. Hence, $\always\psi_1\in q_k\da_i$.\\[2mm]
Finally, assume that $\psi=\comm{j}{\psi_1}$. If $k=0$ then, by definition of initial state and of the  local satisfaction relation, $\psi\notin q_0\da_i$ and $\mu^{\tau}_i,\emptyset\not\sat_i\psi$ and the result follows. If $k>0$ then $\last(\xi^k|_i)=e_k$. Assume first that $\comm{j}{\psi_1}\in q_k\da_i$. Furthermore, let $k_1$ be the greatest index less that $k$ such that $a_{k_1}\cap\Val_i\neq\emptyset$. Then $e_{k_1+1}\in \Ev_i$. We have to consider two cases: either (i) $k_1=k-1$; or (i) $k_1<k-1$. Let us consider case (i). Then, by condition~\ref{dnba:sync1}, it follows that $a_{k_1}\cap\Val_j\neq\emptyset$ and thus $e_k\in\Ev_j$. Furthermore, by condition~\ref{dnba:lc}, it follows that $\psi_i\in q_k\da_j$. Hence, by induction hypothesis, $\mu^\tau_j,\xi^k|_j\sat_j\psi_1$. But, as $\last(\xi^k|_i)=e_k=\last(\xi^k|_j)$ then $\last(\xi^k|_i)\da_j=\last(\xi^k|_j)\da_j=\xi^k|_j$ and so $\mu^{\tau}_j,\last(\xi^k|_i)\da_j\sat_j\psi_1$ which implies that $\mu^{\tau}_i,\xi^k|_i\sat_i\comm{j}{\psi_1}$. Consider now the case (ii). In this case, we know that $\last(\xi^k|_i)=e_{k_1+1}$. We also know that $q_{k_1+1}\da_i=\dots=q_k\da_i$. Hence $\comm{j}{\psi_1}\in q_{k_1+1}\da_i$ and, by condition~\ref{dnba:lc}, it also follows that $\psi_1\in q_{k_1+1}\da_j$. Additionally, by condition~\ref{dnba:sync1}, $a_{k_1}\cap\Val_j\neq\emptyset$, which implies that $e_{k_1+1}\in\Ev_j$. Reasoning as in case (i), we can conclude that $\mu^{\tau}_i,\xi^{k_1+1}|_i\sat_i\comm{j}{\psi_1}$. But $\xi^{k_1+1}|_i=\xi^{k}|_i$ given that $\last(\xi^{k}|_i)=e_{k_1+1}=\last(\xi^{k_1+1}|_i)$. Hence, 
$\mu^{\tau}_i,\xi^{k}|_i\sat_i\comm{j}{\psi_1}$. To prove the converse, assume that $\mu^{\tau}_i,\xi^k|_i\sat_i\comm{j}{\psi_1}$. Then $\last(\xi^k|_i)\in\Ev_j$ and 
$\mu^{\tau}_j,\last(\xi^k|_i)\da_j\sat_j\psi_1$. Let $\last(\xi^k|_i)=e_{k_1}\in\Ev_i$. Clearly, $k_1\leq k$. Again, we need to consider two cases: either (i) $k_1=k$; or (ii) $k_1<k$. In the first case, $e_k\in\Ev_i$ implies that $a_{k-1}\cap\Val_i\neq\emptyset$, $e_k\in\Ev_j$ implies that $a_{k-1}\cap\Val_j\neq\emptyset$ and, by induction hypothesis, $\psi_1\in q_k\da_j$, given that 
$\last(\xi^k|_i)=e_k=\last(\xi^k|_j)$ and so $\mu^{\tau}_j,\last(\xi^k|_i)\da_j\sat_j\psi_1$ implies $\mu^{\tau}_j,\xi^k|_j\sat_j\psi_1$. Hence, using condition~\ref{dnba:sync2}, we conclude that $\comm{j}{\psi_1}\in q_k\da_i$. If condition (ii) holds then 
$a_{k_1-1}\cap\Val_i\neq\emptyset$ and $a_{k_1}\cap\Val_i=\dots=a_{k-1}\cap\Val_i=\emptyset$, which implies that $q_{k_1}\da_i=\dots=q_k\da_i$. Furthermore, given  that $\last((\xi^k|i)\da_j)=e_{k_1}$ then $(\xi^k|i)\da_j=\xi^{k_1}\da_j$. Thus, $\mu^{\tau}_j,\xi^{k_1}\da_j\sat_j\psi_1$ and, by induction hypothesis, $\psi_1\in q_{k_1}\da_j$. Furthermore, like in case (i), we also know that $a_{k_1-1}\cap\Val_i\neq\emptyset$ and $a_{k_1-1}\cap\Val_j\neq\emptyset$. Hence, $\comm{j}{\psi_1}\in q_{k_1}\da_i$, which implies that $\comm{j}{\psi_1}\in q_{k}\da_i$. 
\\[2mm]
Next, we prove a similar result for the global level. For each $\alpha_1\in\closure(\alpha)$, $\alpha_1\in q_0$ iff $\mu^\tau,\xi_0\sat\alpha_1$. We abuse notation and write $\alpha_1\in q_0$ to mean that $\alpha_1\in q_k\da_i$, for some $i\in\Id$. Clearly, $q_0\da_i\cap\subf(\alpha)=q_0\da_j\cap\subf(\alpha)$, that is, the initial states of all the local automata have exactly the same global subformulas of $\alpha$. Again, the proof follows by induction in the structure of $\alpha$. The propositional cases are immediate consequences of the properties of elementary sets. So, let $\alpha_1=\gf{i}{\varphi}$. Then, $\mu^\tau,\xi_0\sat\gf{i}{\varphi}$ iff
$\mu^\tau_i,\xi_0|_i\sat_i\varphi$ iff $\varphi\in q_0\da_i$, by the previous result, iff $\gf{i}{\varphi}\in q_0\da_i$, by the properties of elementary sets, iff $\gf{i}{\varphi}\in q_0\da_i$.
}

We now prove the converse, i.e, we prove that any DTL model of $\alpha$ can be captured by $\cD_\alpha$. Let $\mu=\tuple{\lambda,\val}$ be an interpretation structure and let
$\tuple{\Ev,\leq_{\Ev}}$ be the underlying global order on events. It is
always possible to linearize $\tuple{\Ev,\leq_{\Ev}}$, i.e., it is always
possible to define a bijection $\ell:\nats_1\to\Ev$ such that if
$k_1<_{\nats} k_2$ then $\ell(k_1) <_{\Ev}\ell(k_2)$, where $<_{\nats}$
is the usual ordering on the natural numbers. See, e.g.,
\cite{best:fern:88}. From now on, we assume fixed a linearization function
(or just \emph{linearization}) $\ell$, which induces an enumeration of the global states as follows
\begin{itemize}
\item $\xi^0=\emptyset$;
\item $\xi^{k}=\xi^{k-1}\cup\{\ell(k)\}$ for each $k\geq 1$.
\end{itemize}
Consider the word $w^{\mu,\ell}=a_0a_1a_2\dots$ where
$$
a_k=\bigcup_{i\in\Ids(\ell(k+1))}\{p\in\lit_i\mid \mu_i,\xi^k|_i\sat_i p\}
$$
This word represent one possible \textit{evolution} of the system represented by $\mu$. Our goal is to show that $w^{\mu,\ell}$ is captured by $\cD_\alpha$, that is, to show that $w^{\mu,\ell}\in L(\cD_\alpha)$. 

\begin{theorem}
If $\mu\in\Mod(\alpha)$ then $w^{\mu,\ell}\in L(\cD_\alpha)$.
\end{theorem}
\proof{To show that $w^{\mu,\ell}\in L(A_\alpha)$ we need to present an accepting run for $w^{\mu,\ell}$. For each $k\in\nats$ consider the sets of formulas induced by $\mu$:
\begin{itemize}
\item $x^i_0=\{\alpha\in\closure(\alpha)\mid\mu,\xi^0\sat\alpha\}$;

\item $x^i_{k}=\begin{cases}
				\{\alpha\in\closure(\alpha)\mid\mu,\xi^k\sat\alpha\}&\textrm{if }\last(\xi^k)\in\Ev_i\\
				x^i_{k-1}&\textrm{otherwise}
			\end{cases}$

\item $y_{k}^i=\{\varphi\in\closure(\alpha)\cap\LL_i\mid\mu_i,\xi^k|_i\sat_i\varphi\}$.

\item $q_k=\otimes_{i\in\Id}(x^i_k\cup y_k^i)$.
\end{itemize}
Each $x^i$ has information about the global formulas and each $y^i$ has information about the local formulas of agent $i$. We start by establishing a structural result on the local component of the states. If, for $ i\in \Id$, $\ell(k+1)\notin \Ev_i$ then  $y^i_{k+1}=y^i_{k}$. This is an immediate consequence of the fact that if $\ell(k+1)\notin\Ev_i$ then $\xi^{k+1}|_i=\xi^k|_i$ and so $y^i_{k+1}=y^i_{k}$.

Having established this result, we prove that each $q_k$ is a state in $\cD_{\alpha}$. By construction, we have that $(x^i_k\cup y^i_k)\da_i=x^i_k\cup y_k^i$ and 
$x^i_k\cup y_k^i\subseteq\closure(\alpha)$. Furthermore,  each $x^i_k\cup y_k^i$ is $i$-elementary. The conditions concerning the connectives and temporal operators are a consequence of the definition of the satisfaction relation. The only condition worth checking is the one related to global formulas. In this case, $\gf{i}{\varphi}\in x^i_k\cup y^i_k$ iff $\gf{i}{\varphi}\in x^i_k$ iff $\mu,\xi^k\sat\gf{i}{\varphi}$ iff
$\mu_i,\xi^k|_i\sat_i\varphi$ iff $\varphi\in y^i_k$ iff $\varphi\in x^i_k\cup y^i_k$.

Next, we prove that $q_0q_1q_2\dots $ is a run for $w^{\mu,\ell}$ in $\cD_{\alpha}$, that is, we need to establish that $q_0\in Q_0$ and $q_k\transact{a_k}q_{k+1}$, for every $k\in\nats$. The fact that $q_0\in Q_0$ is straightforward. Indeed, $q_0\in Q_0$ iff $q_0\da_i\in Q_{0_i}$ iff $\alpha\in q_0\da_i$ and $q_0\da_i$ has no communication formulas. Observe that $\alpha\in x^i_0$ because $\mu,\xi^0\sat\alpha$ given that $\mu\in\Mod(\alpha)$. And 
$q_0\da_i$ has no communication formulas because $\xi^0=\emptyset$ and so it does not satisfy any communication formula. To prove that $q_k\transact{a_k}q_{k+1}$ we consider two cases: (i) $a_k\cap\Val_i=\emptyset$; and (ii) $a_k\cap\Val_i\neq\emptyset$, for  $i\in\Id$.\\[1mm]
(i) If $a_k\cap\Val_i=\emptyset$ then $i\notin\Ids(\ell(k+1))$ which implies that $y^i_{k+1}=y^i_{k}$. Furthermore, by definition, $x^i_{k+1}=x^i_{k}$. Hence, $q_{k+1}\da_i=q_k\da_i$. \\[1mm]
(ii) If $a_k\cap\Val_i\neq\emptyset$ then $\ell(k+1)\in\Ev_i$ and $\xi^{k+1}|_i=\xi^{k}|_i\cup\{\ell(k+1)\}$. We prove that, in this case, $q_{k+1}\da_i\in\delta_i(q_k\da_i,a_k\cap\Val_i)$:
\begin{itemize}
\item $a_k\cap\Val_i=q_k\da_i\cap\Val_i$: straightforward by construction of $a_k$ and $q_k$;

\item let $\nxt\psi\in \closure(\alpha)\cap \LL_i$: $\nxt\psi\in q_{k}\da_i$ iff $\nxt\psi\in y^i_{k}$ iff $\mu_i,\xi^k|_i\sat_i\nxt\psi$ iff
$\mu_i,\xi^{k}|_i\cup\{\ell(k+1)\}\sat_i\psi$ iff $\mu_i,\xi^{k+1}|_i\sat_i\psi$ iff $\psi\in y^i_{k+1}\da_i$ iff $\psi\in q_{k+1}\da_i$;

\item let $\always\psi\in \closure(\alpha)\cap \LL_i$: $\always\psi\in q_{k}\da_i$ iff $\always\psi\in y^i_{k}$ iff $\mu_i,\xi^k|_i\sat_i\always\psi$ iff
$\mu_i,\xi'\sat_i\psi$, for $\xi'\supseteq\xi^{k}|_i$, iff $\mu_i,\xi^k|_i\sat_i\psi$ and $\mu_i,\xi'\sat_i\psi$, for $\xi'\supseteq\xi^{k}|_i\cup\{\ell(k+1)\}$, iff $\mu_i,\xi^k|_i\sat_i\psi$ and $\mu_i,\xi^{k}|_i\cup\{\ell(k+1)\}\sat_i\always\psi$ iff $\mu_i,\xi^k|_i\sat_i\psi$ and $\mu_i,\xi^{k+1}|_i\sat_i\always\psi$ iff $\psi\in y^i_k$ and $\always\psi\in y^i_{k+1}$ iff $\psi\in q_k\da_i$ and $\always\psi\in q_{k+1}\da_i$.
\end{itemize}
Hence, from (i) and (ii),  if follows that $q_{k+1}\in\delta(a_k,q_k)$, as desired.

Finally, we just need to establish that the run is accepting. We start by showing that each local run is accepting. Let $\always\varphi\in \closure(\alpha)\cap\LL_i$. We need to show that there are infinitely many indices $k$ such that $q_k\da_i\in F_{\always\varphi}$. Assume that this is not the case, i.e., assume that there are only finitely many indices $k$ such that $q_k\da_i\in F_{\always\varphi}$. Then, there is $n\in\nats$ such that $q_k\da_i\notin  F_{\always\varphi}$, for every $k\geq n$. But, if $q_k\da_i\notin  F_{\always\varphi}$ then $\always\varphi\notin q_k\da_i$ and $\varphi\in q_k\da_i$, for $k\geq n$. If $\always\varphi\notin q_k\da_i$ then $\always\varphi\notin y^i_k$ and so
$\mu_i,\xi^k|_i\not\sat\always\varphi$. This implies that there is $\xi'\supseteq\xi^k|_i$ such that $\mu_i,\xi'\not\sat\varphi$. It is not very difficult to see that there is $m\geq k$ such that $\xi'=\xi^m|_i$. Hence $\mu_i,\xi^m|_i\not\sat\varphi$ which implies that $\varphi\notin y^i_m$ and, consequently, $\varphi\notin q_m\da_i$. But, as $m\geq k\geq n$ then $\varphi\in q_m\da_i$ and we reach a contradiction. This means that each set $F_{\always\varphi}$ is visited infinitely often and, thus, the local run is accepting. The fact that the global word $w^{\mu,\ell}$ is fair is a consequence of the linearization. Then, by Lemma~\ref{dnba:localvsglobal}, we can conclude that the global run is also accepting.
}

We now study the relationship between $\mu$ and $\mu^{\tau}$ where $\tau$ is an accepting run for $w^{\mu,\ell}$, and between $w$ and $w^{\mu^\tau,\ell}$.
If we start  with a model $\mu$, construct $w^{\mu,\ell}$ and then define $\mu^{w^{\mu,\ell}}$, we may wonder what is the relation between $\mu$ and $\mu^{w^{\mu,\ell}}$? Conversely, if we start with a word $w$ (with accepting run $\tau$) and we build the model $\mu^\tau$, can we can find a linearization $\ell$ such that $w^{\mu^\tau,\ell}$ is $w$? The following lemmas answer both these questions.

\begin{lemma}\label{lemma:iso1}
Let $w\in L(\cD_{\alpha})$ with accepting run $\tau$. Then, there is a linearization $\ell$ such that $w=w^{\mu^\tau,\ell}$.
\end{lemma}
\proof{
Let $w\in L(\cD_{\alpha})$ with accepting run $\tau$ and consider $\mu^{\tau}$ defined as above. Consider the linearization $\ell:\nats_1\to \Ev$ such that $\ell(k)=e_k$.  First, we observe that $\xi^0=\emptyset$ and $\xi^k=\{\ell(1),\dots,\ell(k)\}=\{e_1,\dots,e_k\}$. Let $p\in \Prop_i$. Then, for $k\in \nats$
$$
\begin{array}{rcll}
p\in w^{\mu^\tau,\ell}_k & \textrm{iff}	& \mu^{\tau}_i,\xi^k|_i\sat_i p & \textrm{(by definition of $w^{\mu,\ell}$)}\\
						 & \textrm{iff} & p\in\vartheta_i(\xi^k|_i)   & \textrm{(by definition of satisfaction)}\\
						 & \textrm{iff} & p\in q_k\da_i               & \textrm{(by definition of $\mu^{\tau}$)}\\
						 & \textrm{iff} & p\in w_k\cap\Val_i   & \textrm{(by definition of $\delta_i$)}\\
						 & \textrm{iff}  & p\in w_k 					& \textrm{(because $p\in\Prop_i$)}.
\end{array}
$$
Hence, $w^{\mu^\tau,\ell}_k=w_k$, for every $k\in\nats$, and so $w^{\mu^\tau,\ell}=w$.
}

We can state a similar result for $\mu$ and $\mu^{w^{\mu,\ell}}$. But in case, we cannot say that the two interpretation structures are equal but only isomorphic. We say that two distributed life-cycles $\lambda_1=\{\tuple{\Ev^1_i,\leq^1_i}_i\}_{i\in\Id}$ and $\lambda_2=\{\tuple{\Ev^2_i,\leq^2_i}_i\}_{i\in\Id}$ are \textit{isomorphic}, written $\lambda_1\cong_f\lambda_2$, if there is an bijection $f:\Ev^1\to \Ev^2$ such that $e\leq^1 e'$ if and only if $f(e)\leq^2f(e')$, for every $e,e'\in\Ev^1$. Function $f$ establishes a bijection between states of $\lambda_1$ and $\lambda_2$  as would be expected.
We say that two interpretation structures $\mu_1=\tuple{\lambda_1,\vartheta_1}$ and $\mu_2=\tuple{\lambda_2,\vartheta_2}$ are \textit{isomorphic}, written $\mu_1\cong_f\mu_2$, if $\lambda_1\cong_f\lambda_2$ and $\vartheta_i^1(\xi_i)=\vartheta_i^2(f(\xi_i))$, for every state $\xi_i\in \Xi^1_i$. We may drop the reference to $f$ and simply write $\lambda_1\cong\lambda_2$.  We can now state the converse result of Lemma~\ref{lemma:iso1}.

\begin{lemma}\label{lemma:iso2}
Let $\mu\in\Mod(\alpha)$. Then,  $\mu\cong\mu^{\tau}$, where $\tau$ is an accepting run for $w^{\mu,\ell}$, for a given linearization $\ell$.
\end{lemma}
\proof{
We start by defining a bijection between $\Ev$ and $\Ev^\tau$, where $\Ev^\tau$ denotes the set of events of $\mu^\tau$. Let $f:\Ev\to\Ev^\tau$ be such that $f(e)=e_k$ where $k$ is such that $\ell(k)=e$ that exists because $\ell$ is a linearization of $\Ev$. Furthermore, let $e,e'\in\Ev$ such that $\ell(k)=e$ and $\ell(k')=e'$, for some $k,k'\in\nats_1$. Then $e\leq e'$ iff $\ell(k)\leq\ell(k')$ iff $k\leq k'$ iff $e_k\leq^\tau e_{k'}$. Hence, $\lambda\cong\lambda^\tau$. As it was said before, $f(\emptyset)=\emptyset$ and $f(\xi^k)=f(\{\ell(1),\dots,\ell(k)\})=\{e_1,\dots,e_k\}$. Next, we prove that  $\mu\cong\mu^{\tau}$.
Let $p\in \Prop_i$. Then for $k\in \nats$
$$
\begin{array}{rcll}
p\in \vartheta^{\tau}_i(f(\xi^k)|_i) & \textrm{iff}	& p\in \vartheta^{\tau}_i(\{e_1,\dots,e_k\}|_i) &\textrm{(as observed above)}\\
									& \textrm{iff}	& p\in q_k\da_i                                & \textrm{(by definition of $\mu^{\tau}$)}\\
									& \textrm{iff}	& p\in w^{\mu^\tau,\ell}_k\cap\Val_i & \textrm{(by definition of $\delta_i$)}\\
									& \textrm{iff}	& p\in w^{\mu^\tau,\ell}_k   					& \textrm{(because $p\in\Prop_i$)}\\
									& \textrm{iff} & \mu^{\tau}_i,\xi^k|_i\sat_i p & \textrm{(by definition of $w^{\mu,\ell}$)}\\
								    & \textrm{iff} & p\in\vartheta_i(\xi^k|_i)   & \textrm{(by definition of satisfaction)}.
\end{array}
$$
Hence, we conclude that $\mu\cong\mu^{\tau}$.
}

These two lemmas allow us to conclude that $\Mod(\alpha)$ and $L(D_\alpha)$ have essentially the same information.

\section{Concluding remarks}\label{sec:conc}

We have proposed a notion of \textit{distributed B\"uchi automaton}. We have then endowed DTL with an operation 
semantics based on DNBA's, where for the local components (that have a linear behaviour) an approach similar to 
the followed in~\cite{var:wol:94,bai:kat:08} was adopted. The construction was proved correct with respect to DTL semantics.

As future work, we believe that it would be interesting to extend our approach to other temporal operators, like the until operator and past operators. No surprises are expected as these have been widely studied for LTL and the local agents of DTL have a linear time behaviour.

The main goal of this ongoing work is to endow DTL with a model-checking tool. The work presented in this paper is the first step towards that goal. Having such a tool will allows us to verify some of the problems to which DTL has successfully been applied, but in an automated way~\cite{bas:ccal:jabr:vig:10,ccal:vig:bas:04b,ccal:vig:bas:04c}. It is also our goal to study the complexity of 
our intendend approach and compare it with existing tools. 

\section*{Acknowledgements}
This work was supported by the Portuguese Funda\c c\~ao para a Ci\^encia e a Tecnologia (FCT) by
way of grant UID/EEA/50008/2013 to Instituto de Telecomunica\c c\~{o}es (IT).

\bibliographystyle{plain}
\bibliography{dtl-BuchiAutomata}
\end{document}